\documentclass{PoS}

\title{Probing quantum gravity using high-energy astrophysics}

\ShortTitle{Planck stars}

\author{\speaker{Justine Tarrant}, Geoff Beck and Sergio Colafrancesco\thanks{In memory of Professor Sergio Colafrancesco who sadly passed away before this work was completed.}\\
       School of Physics, University of the Witwatersrand, Private Bag 3, WITS-2050, Johannesburg, South Africa\\
        E-mail: \email{justine.tarrant@wits.co.za}}

\abstract{
	The events observed by LIGO indicate the existence of a large population of intermediate mass black holes. This unexpected result lead to a resurgence in the interest in theories of the formation of primordial black holes with several studies showing that broad mass distributions can evade or satisfy the stringent constraints on monochromatic populations. If such large populations exist they provide the perfect test-bed for theories beyond the standard model of physics that modify black hole evolution. The case we studied is that of "Planck stars", a hypothetical modification of the black hole evolution where it explodes via quantum loop gravity motivated tunnelling. We determine what the high-frequency background signal of such objects exploding over the whole of cosmic history would look like for various black hole populations to place actual empirical constraints on quantum loop gravity via comparison to observed isotropic background signals at the same frequencies. We find that stringent constraints heavily restrict the amount of energy released via the high-energy channel, thereby casting doubt on whether or not the high-energy signal could result in gamma-ray bursts as speculated in the literature.}

\FullConference{High Energy Astrophysics in Southern Africa - HEASA2018\\
		1-3 August, 2018\\
		Parys, Free State, South Africa}

\begin{document}

\section{Introduction}

Building a consistent quantum theory of gravity is challenging, especially when it comes down to verification \cite{Miao:2019pxw,rovelli2014}. This is because it is very difficult to observe the quantum effects of gravity due to the conjectured scale being Planckian \cite{rovelli2014,Liberati:2011bp,Hossenfelder:2010zj}. Therefore, finding ways to probe quantum gravity, perhaps at scales larger than the Planckian scale, would be beneficial to modern physics. Planck stars provide such an avenue for exploration. Planck stars are a recent phenomenon \cite{planck2014,rovelli2014,rovelli2017}, which represent an alternate fate for the death of a black hole. They form when a collapsing shell of matter reaches the Planck density, roughly equivalent to the concerned mass being compressed into a volumetric size near that of a proton. Then, according to loop quantum gravity calculations \cite{hawking1975}, it stops collapsing and `bounces' due to a quantum pressure not unlike the pressure preventing electrons from falling onto an atomic nucleus \cite{rovelli2017}. 

Black holes are known to be stable in classical general relativity. However, they may decay via Hawking radiation \cite{Hawking:1974rv} - a process that takes up to $10^{50}$ Hubble times for a stellar mass black hole \cite{rovelli2017}. As a result, Hawking radiation is not a cosmologically significant mechanism. A Planck star, which is effectively a black hole tunnelling into a while hole and exploding through its event horizon \cite{barrau2014b}, takes place much faster. The bounce and tunnelling occur very rapidly in the rest-frame of the bouncing black hole \cite{rovelli2014}, but due to the large relativistic time dilation, the bounce appears to take place on the order of a billion years for black holes larger than $10^{-7} M_{\odot}$ \cite{barrau2014b,barrau2014}. Black holes larger than  $10^{-7} M_{\odot}$ would not be exploding in the present epoch, therefore if we are to observe such an event we would require the presence of primordial black holes since they are much smaller than a solar mass.

The formation of primordial black holes in the very early universe has been widely discussed \cite{pbh1}. Due to their size they interact weakly with other massive bodies and have therefore been suggested as dark matter candidates \cite{pbh2}. Whilst early research considered primordial black holes which were smaller than a solar mass, stringent constraints~\cite{capela2013} prevent these primordial black holes from making up a large fraction of dark matter. More recent work has revealed that using extended mass distributions may weaken the constraints placed by monochromatic mass functions. The constraints on the dark matter fraction constituted by primordial black holes lie across a wide range of possible primordial black hole masses (see the references in \cite{bellomo2017,kuhnel2017} for example). These come from diverse sources such as microlensing of stars, femtolensing of gamma-ray bursts, and extra-galactic gamma-ray emissions from black hole evaporation~\cite{Carr:2009jm,Griest:2013aaa,2012PhRvD..86d3001B,Tisserand:2006zx}.

We consider two cases that both assume lognormal mass distributions: that of \cite{kuhnel2017} (mean mass $\sim 10^{-9}$ $M_{\odot}$) where the relevant limits arise from Kepler microlensing data~\cite{Griest:2013aaa}, this mass range being chosen because the constraints are less severe than anywhere else in the parameter space. In the case of the second work \cite{bellomo2017} a LIGO relevant mass range is considered instead (mean mass $\sim 10$ $M_{\odot}$). The constraints from  millilensing of quasars limit the primordial black holes to constitute $10$\% of cosmological dark matter in this mass range when monochromatic mass distributions are assumed~\cite{PhysRevLett.86.584}. However, \cite{bellomo2017} shows that if an extended mass distribution is used instead, then these constraints are relaxed sufficiently to allow the primordial black holes to account for practically all cosmological dark matter. This has, when combined with LIGO~\cite{mpbh2}, renewed interest in more massive primordial black holes \cite{mpbh1,mpbh3,carr2018}.
The use of the lognormal mass distributions is important as they have been shown to approximately model the seeding of primordial black hole density perturbations by a number of inflationary scenarios \cite{inf1,inf2,inf3}, allowing us to be as general as possible. 

Finally, we note that pulsar timing arrays also place constraints on the abundance of dark matter in the form of primordial black holes. Using 143 known millisecond pulsars and a 30 year observation time yields $f_{pbh}\sim 0.09$ for $1<M<10M_{\odot}$, but $f_{pbh}\sim 0.5$ in the LIGO window ($\sim 30M_{\odot}$) \cite{Schutz:2016khr}. However, these constraints are strengthened with the implementation of the 2000 pulsars making up the Square Kilometre Array. In the same amount of observation time, SKA places $f_{pbh}\sim 0.05$, ten times smaller than current pulsar timing arrays. Therefore the SKA will be an invaluable resource to constrain the primordial black hole population in the future.

In this paper we discuss the possible high-energy emission coming from Planck stars exploding since the epoch of matter-radiation equality. We consider lognormal mass distributions which both evade and satisfy constraints, as mentioned already.  Furthermore, we compare the spectra obtained (see figure 1) to the isotropic gamma-ray background \cite{Ackermann:2014usa}. This comparison allows one to set a limit on the amount of energy going into gamma-rays as a result of the Planck star explosion. We also compare the results herein to earlier work on the low-energy emission. In both cases we find that a stringent constraint is placed on the amount of energy released in either radio- or gamma-rays, making the Planck star mechanism look less likely to account for fast radio bursts or gamma-ray bursts, as speculated to in the literature.

This paper is organized as follows: section 2 gives a layout of the formalism used for calculating the high-energy spectrum of the Planck stars exploding up until the present epoch. Section 3 presents the data used for comparison with spectra. In section 4 we discuss the results and conclude in section 5.

\section{High-energy emission}

The flux for the high-frequency background radiation generated by exploding Planck stars, from the time of matter-radiation equality till the present epoch, is given by
\begin{equation}
\Phi (\lambda) = \int_{z_{eq}}^{0} dz \, \frac{dV}{dz} N(z)\frac{c}{2 R_{BH}(z)}\frac{\chi(\lambda,z)}{4 \pi D_L^2} \; ,
\end{equation}
which is built up as follows: $\frac{dV}{dz}$ is the differential co-moving volume element, $\frac{2 R_{BH}(z)}{c}$ is the assumed time taken for the emission of energy from the primordial black hole exploding at $z$, i.e. the time taken to travel across the diameter of the black hole, $N(z)$ is the number density of exploding planck stars at each redshift $z$, $\chi(\lambda,z)$ is the spectral energy distribution of an individual explosion and $D_L$ is the co-moving distance at redshift $z$. The above flux may also be used for the accompanying low-energy signal, however the form of the spectral energy distribution $\chi$ will be different.

We may calculate $N(z)$ by noticing the following:
\begin{equation}
N(M)dM = N(\tau)d\tau = N(z)dz\; .
\end{equation}

\noindent where 

\begin{equation}
N(z) = N(M)\frac{dM}{d\tau}\frac{d\tau}{dz}. 
\end{equation}

The term $\frac{dM}{d\tau}$ is found through the following relationship~\cite{rovelli2014}
\begin{equation}
\tau = \left(\frac{M}{M_{pl}}\right)^2t_{pl}
\end{equation}
which relates the lifetime of the black hole $\tau$ to its mass $M$. Here $t_{pl}$ and $M_{pl}$ are the Planck time and mass, respectively. $N(M)$ is the number density of Planck stars per unit mass. In this case we choose $N(M)$ to be lognormal, following both \cite{bellomo2017,kuhnel2017}. The lognormal distribution is convenient as it fits a wide range of inflationary models for primordial black holes~\cite{inf1,inf2,inf3}. Furthermore, we normalise $N(M)$ to some fraction $f_{pbh}$ of the total dark matter density being composed of primordial black holes. This is because we know how much dark matter there is in the universe and since primordial black holes are weakly interacting, they are considered dark matter candidates. Here we treat Planck stars as being made of exploding primordial black holes, therefore the last assumption means we may estimate the number of Planck stars in the universe.

The shape of the spectral energy distribution $\chi$ follows a thermal distribution. This is because the photon gas emitted upon explosion was captured in the hot early universe. One may normalise the spectral energy distribution in two ways. Firstly, we assume that $\chi$ has a thermal shape normalised to some fraction $\chi_0$ of the total mass-energy of the black hole and peaked at $\lambda_{obs}$, given by~\cite{barrau2014b} 

\begin{equation}
\lambda_{obs} \sim \lambda_{em} (1+z) \left[ \sinh^{-1}{\left({\sqrt{\frac{\Omega_{\Lambda}}{\Omega_m}}}(z+1)^{-3/2}\right)}\right]^{1/4} \; .
\end{equation}

\noindent where $\lambda_{obs}$ and $\lambda_{em}$ are the observed and emitted wavelengths, respectively. $\Omega_m$ and $\Omega_{\Lambda}$ are the matter and cosmological constant density parameters, respectively~ \cite{planck2014}. The first approach contains the two free parameters $f_{pbh}$ and $\chi_0$. Secondly, we may choose to normalise $\chi$ using the temperature $T\propto (hc)/(k_B\lambda_{obs})$, which is also peaked at $\lambda_{obs}$, and has the single parameter $f_{pbh}$. The latter will directly limit $f_{pbh}$, whilst the former option will allow for more freedom as $\chi_0$ shifts around in the parameter space.

\section{Isotropic gamma ray background}

Our gamma-ray background data is sourced from \cite{Ackermann:2014usa}. We then compare this data to the predicted high-energy background emission caused by Planck star explosions. We expect the Planck star emissions over all epochs to contribute an extra component to the isotropic high-energy backgrounds, and here we predict how large that component should be using the free parameters $f_{pbh}$ and $\chi_0$. Furthermore, any value of the product $f_{pbh}\chi_0$ that allows the Planck star spectrum to exceed the isotropic gamma-ray background by a 3$\sigma$ confidence level or more will be excluded. This is analogous to the treatment of the low-energy signal where we compared the predicted spectrum to the extra-galactic background light.

\section{Results}

Figure 1 shows the spectra for exploding Planck stars originating from primordial black holes under the assumption that primordial black holes may constitute the majority of dark matter ($f_{pbh}\sim 1$). Both spectra are built using lognormal mass distributions $N(M)$. For the case where $\mu=10 M_{\odot}$ and $\sigma = 0.25$ \cite{bellomo2017}, a minimal $\chi^2$ fitting yields $\chi_0 f_{pbh}\leq 10^{-38}$. For $f_{pbh}\sim 1$, we see that $\chi_0 \leq 10^{-38}$ and for a black hole exploding today ($\sim 10^{-7}M_{\odot}$) roughly $\sim 10^{47}$erg is released. Therefore, the amount of energy going into gamma-rays during the explosion is stringently constrained to $\sim 10^9$erg. Similarly, for $\mu=10^{-9} M_{\odot}$ and $\sigma = 0.5$ \cite{kuhnel2017} we find $\chi_0 f_{pbh}\leq 10^{-39}$, resulting in a $\sim 10^8$erg energy release, thereby also placing a stringent constraint on the energy going into gamma-rays. These energies are far too small to be considered as gamma-ray bursts.

Furthermore, suppose that instead a large portion of the energy from the Planck star explosion \textit{did} go into producing gamma-rays, i.e. take $\chi_0\sim10^{-9}$ and  $f_{pbh} \sim 10^{-30}$ in which case the number of Planck stars producing gamma-ray bursts would be very small, almost negligible. But, when they do explode they will result in a release of $10^{38}$erg in energy.

The previous cases dealt with spectral energy distribution which was Planckian, where we normalise using the $\chi_0$ parameter. When choosing to fix the normalisation according to the temperature, we find that $f_{pbh}\leq 10^{-30}$. Thus, very few Planck stars would result in gamma-ray bursts. When we integrate over the thermal distribution to extract an energy for the gamma-rays resulting from an explosion of a $10^{-7} M_{\odot}$ black hole we find that $10^{38}$erg is released, as we had above. Therefore to get a reasonable amount of energy out, we need to severely reduce the primordial black hole population producing Planck stars, regardless of the spectral normalisation used.

For the low-energy signal we found that similar amounts of energy would be released, i.e. $\chi_0 f_{pbh} \leq 10^{-34}$ \cite{Tarrant:2019tgv}. Therefore, making it unlikely that Planck stars could result in fast radio bursts.

Our results show some tension with the LIGO observations as they imply that less than one $10M_{\odot}$ black hole would be present in the entire Milky Way, which is hard to reconcile with the observed merger rate inferred by LIGO. This tension is released if all LIGO black holes were of astrophysical origin rather than primordial. Therefore, if the Planck star hypothesis is true then LIGO black holes can't be primordial.

\section{Conclusion}

Probing quantum gravity is a non-trivial task. However, if objects like Planck stars exist, there may be a window onto quantum gravity after all. Two main sigmals are expected: a high- and low-energy component. And it is suggested in the literature that the former may result in fast radio bursts whilst the latter may result in gamma-ray bursts. In this article we consider the high-energy channel, but note that the low-energy signal has a similar fate: using the models herein we show that it is unlikely that Planck stars result in fast radio bursts or gamma-ray bursts.

Using the isotropic gamma-ray background, we were able to show that, for the models discussed, we can severely restrict the total energy emitted as gamma-rays from a Planck star explosion in the present epoch to be $\leq 10^{9}$erg when considering $f_{pbh}\sim 1$. Considering the case where a large amount of energy does go into gamma-rays comes at a cost. Namely, the population of primordial black holes producing Planck stars is restricted via stringent constraints on $f_{pbh}$, making the events very rare.

We also note that despite  using two mass function models, i.e. Bellomo and Kuhnel, which vary greatly in terms of their mean mass, only one order of magnitude  separates their limits on the Planck star high-energy emissions, i.e. $10^{9}$ vs $10^8$erg. Furthermore, the Planck spectral energy distribution and the thermal distribution normalized using the temperature are consistent with each other, in that under the same circumstances for the $\chi_0$ and $f_{pbh}$, we find the results are the same. Therefore, the given insensitivity to changes in the mean mass and the consistency of the two spectral energy distributions suggests that the result is robust.

Therefore, the constraints obtained cast doubt upon whether or not the Planck stars described using the models discussed here could account for gamma-ray bursts as speculated to in the literature since either a tiny population of primordial black holes is required (making their occurrence very rare), or one places stringent constraints the total energy.

\section*{Acknowledgements} 

\noindent This work is based on the research supported by the South African
Research Chairs Initiative of the Department of Science and Technology
and National Research Foundation of South Africa (Grant No 77948).

\begin{figure}[htbp]
	\centering
	\resizebox{0.7\hsize}{!}{\includegraphics{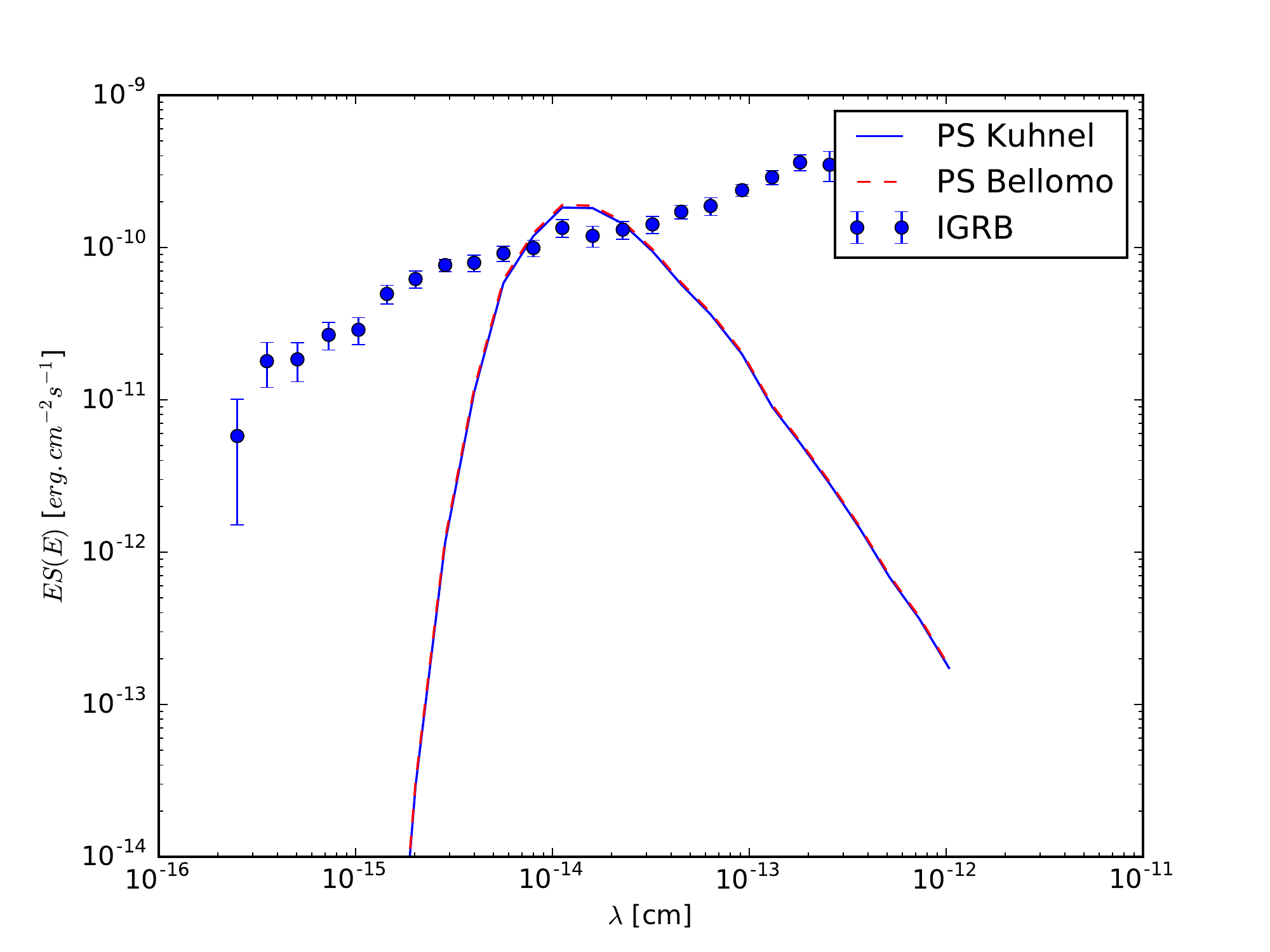}}
	\caption{Comparison of high-frequency Planck star background (PS) to isotropic gamma ray background (IGRB) data from \cite{Ackermann:2014usa}.}
	\label{fig:ps1}
\end{figure}

\bibliographystyle{iopart-num}
\bibliography{planck_stars.bib}

\end{document}